\documentclass[aps,prl,twocolumn,superscriptaddress,showpacs,preprintnumbers,amsmath,amssymb,
longbibliography,
nofootinbib]{revtex4-2}

\usepackage{graphicx}
\graphicspath{{ps}}

\usepackage{orcidlink}

\newcommand{\res}{\ensuremath{0.22\pm0.94\pm0.42}} 
\newcommand{\xip}{\ensuremath{\xi^\prime}}
\newcommand{\ee}{\ensuremath{e^+ e^-}}
\newcommand{\tat}{\ensuremath{\tau^+ \tau^-}}
\newcommand{\qq}{\ensuremath{q\bar{q}}}
\newcommand{\tlnn}{\ensuremath{\tau^- \to \ell^- \bar{\nu}_\ell\nu_\tau}}
\newcommand{\tmnn}{\ensuremath{\tau^- \to \mu^- \bar{\nu}_\mu\nu_\tau}}
\newcommand{\menn}{\ensuremath{\mu^- \to e^- \bar{\nu}_e\nu_\mu}}
\newcommand{\kpipi}{\ensuremath{K^- \to \pi^-\pi^0}}
\newcommand{\kpipipi}{\ensuremath{K^- \to \pi^-\pi^0\pi^0}}
\newcommand{\kcpi}{\ensuremath{K^- \to \pi^-\pi^+\pi^-}}
\newcommand{\kmunu}{\ensuremath{K^- \to \mu^- \bar{\nu}_{\mu}}}
\newcommand{\kpmunu}{\ensuremath{K^- \to \pi^0 \mu^- \bar{\nu}_{\mu}}}
\newcommand{\kpenu}{\ensuremath{K^- \to \pi^0 e^- \bar{\nu}_{e}}}
\newcommand{\pimunu}{\ensuremath{\pi^- \to \mu^- \bar{\nu}_{\mu}}}
\newcommand{\cascade}{\ensuremath{\tau^- \to \mu^- (\to e^- \bar{\nu}_e\nu_\mu) \bar{\nu}_\mu\nu_\tau}}

\begin{document}

\title{\quad \\[0.5cm] {First measurement of the Michel parameter {\boldmath{\xip}} in the {\boldmath{$\tmnn$}} decay at Belle}}

\noaffiliation
  \author{D.~Bodrov\,\orcidlink{0000-0001-5279-4787}} 
  \author{P.~Pakhlov\,\orcidlink{0000-0001-7426-4824}} 
  \author{I.~Adachi\,\orcidlink{0000-0003-2287-0173}} 
  \author{H.~Aihara\,\orcidlink{0000-0002-1907-5964}} 
  \author{S.~Al~Said\,\orcidlink{0000-0002-4895-3869}} 
  \author{D.~M.~Asner\,\orcidlink{0000-0002-1586-5790}} 
  \author{H.~Atmacan\,\orcidlink{0000-0003-2435-501X}} 
  \author{T.~Aushev\,\orcidlink{0000-0002-6347-7055}} 
  \author{R.~Ayad\,\orcidlink{0000-0003-3466-9290}} 
  \author{V.~Babu\,\orcidlink{0000-0003-0419-6912}} 
% ...
\collaboration{The Belle Collaboration}

\begin{abstract}
We report the first measurement of the Michel parameter \xip\ in the $\tmnn$ decay with a new method proposed just recently. The measurement is based on the reconstruction of the $\tmnn$ events with subsequent muon decay-in-flight in the Belle central drift chamber. The analyzed data sample of $988\,\text{fb}^{-1}$ collected by the Belle detector corresponds to approximately $912\times10^6$ $\tat$ pairs. We measure $\xip=0.22\pm0.94(\text{stat})\pm0.42(\text{syst})$, which is in agreement with the standard model prediction of $\xip=1$. Statistical uncertainty dominates in this study, being a limiting factor, while systematic uncertainty is well under control. Our analysis proved the practicability of this promising method and its prospects for further precise measurement in future experiments.
\end{abstract}

\pacs{13.35.Dx, 12.15.Ji, 13.66.De}
\maketitle
\tighten
\renewcommand{\thefootnote}{\fnsymbol{footnote}}
\setcounter{footnote}{0}
In the standard model (SM), the $\tau$ lepton decays via weak charged vector current interaction mediated by a virtual $W$ boson that interacts only with left-handed fermions. Since the momentum transfer in $\tau$ decays is much smaller than the $W$ mass, the decay amplitude can be approximated with high accuracy by a four-fermion interaction with $V-A$ Lorentz structure. Any deviations from this structure will indicate physics beyond the SM caused by either anomalous coupling constants of the $W$--$\tau$ interaction or by a contribution of new gauge or charged Higgs bosons~\cite{Bryman:2021teu}. The larger mass of the $\tau$ lepton compared to that of the muon can enhance the particular new physics contribution to the decay process leading to an ``incorrect'' Lorentz structure~\cite{Krawczyk:2004na,Chun:2016hzs,Marquez:2022bpg}. 

Pure leptonic $\tau$ decays hold a special place in studies of the charged weak interaction since the theoretical calculation of these processes can be done precisely without QCD-associated uncertainties. For the complete study of the Lorentz structure of the interaction responsible for the leptonic $\tau$ decay, one starts with the Lorentz invariant, local, derivative-free, lepton-number-conserving Hamiltonian of the four-fermion interaction~\cite{Michel:1949qe}. The resulting matrix element of the $\tlnn$~\cite{Note1} decay, written in the form of helicity projections~\cite{Scheck:1984md,Mursula:1984zb,Fetscher:1986uj}, includes scalar, vector, and tensor interactions with left- and right-handed initial $\tau$ and daughter leptons. The relative contribution of each term is determined by ten corresponding complex coupling constants. They are normalized so that the general interaction strength is given by the Fermi constant $G_F$. The experimentally observed differential decay width of the lepton is determined by bilinear combinations of coupling constants. It is convenient to express them in terms of the so-called Michel parameters described in detail elsewhere~\cite{Fetscher:1993sxo}.

At present, all Michel parameters have been measured with high precision in $\menn$ decays~\cite{Workman:2022ynf}, while in $\tau$ decays, only four Michel parameters have been measured with an accuracy to the order of a few percent levels~\cite{Workman:2022ynf}.
All measurements are consistent with the predictions of the SM within experimental uncertainties. To obtain the remaining Michel parameters in $\tau$ decays, it is required to measure the daughter lepton polarization, which has not been done directly yet. Recently, the Belle Collaboration reported the measurement of two Michel parameters associated with the daughter lepton polarization in the radiative leptonic $\tau$ decays~\cite{Shimizu:2017dpq}. However, this measurement suffers from too large uncertainties. Five-body leptonic $\tau$ decays open up another opportunity to measure Michel parameters related to the polarization of the daughter lepton from $\tlnn$~\cite{Flores-Tlalpa:2015vga}. Although a sensitivity test for this method was performed by the Belle Collaboration~\cite{Sasaki:2017unu, Sasaki:2017msf}, demonstrating its feasibility, the measurement has not been done yet.

In this Letter, we present the first direct measurement of the Michel parameter $\xip$ in the $\tmnn$ decay. This parameter determines the longitudinal polarization of the daughter muon in the case of an unpolarized $\tau$ lepton. We use the method developed in Ref.~\cite{Bodrov:2022mbd} for implementation in the experiments at the $B$ factories. It is based on the reconstruction of the $\menn$ decay-in-flight in the tracking detector and exploits the correlation between the daughter electron momentum and the muon spin. A more comprehensive description of the measurement is provided in a companion article~\cite{Belle:2023dyc}.

This analysis is based on a data sample collected at center-of-mass energies around the $\Upsilon(nS)$ resonances ($n\in\{1,\,2,\,3,\,4,\,5\}$) with an integrated luminosity of $988\,\text{fb}^{-1}$ corresponding to $\approx912\times 10^6$ \tat\ pairs. We use Monte Carlo (MC) simulation of the main processes in $\ee$ annihilation to optimize the selection criteria, study the background contamination, and determine the fit function.

The data are collected with the Belle detector~\cite{Belle:2000cnh, Belle:2012iwr} at the KEKB asymmetric-energy \ee\ collider~\cite{Kurokawa:2001nw, *[][{ and references therein.}] Abe:2013kxa}. The Belle detector is a large-solid-angle magnetic spectrometer that consists of a silicon vertex detector (SVD), a 50-layer central drift chamber (CDC), an array of aerogel threshold Cherenkov counters (ACC), a barrel-like arrangement of time-of-flight scintillation counters (TOF), and an electromagnetic calorimeter (ECL) composed of CsI(Tl) crystals located inside a superconducting solenoid coil that provides a 1.5~T magnetic field. An iron flux return located outside of the coil is instrumented to detect $K_L^0$ mesons and to identify muons (KLM).

The most critical Belle subdetector for this study is the CDC~\cite{Hirano:2000ei}. It is large enough (the outer radius is $874\,\text{mm}$) to reliably reconstruct both the daughter electron and mother muon tracks if the decay occurs in the inner CDC volume. However, the average flight distance of muons from $\tau$ decays at Belle is of the order of a kilometer, leading to a decay probability inside the CDC of about $10^{-3}$. The large \tat\ sample compensates for this tiny probability and allows one to reconstruct hundreds of such events at Belle.

\begin{figure}[htb]
  \centering
  \includegraphics[width=1\linewidth]{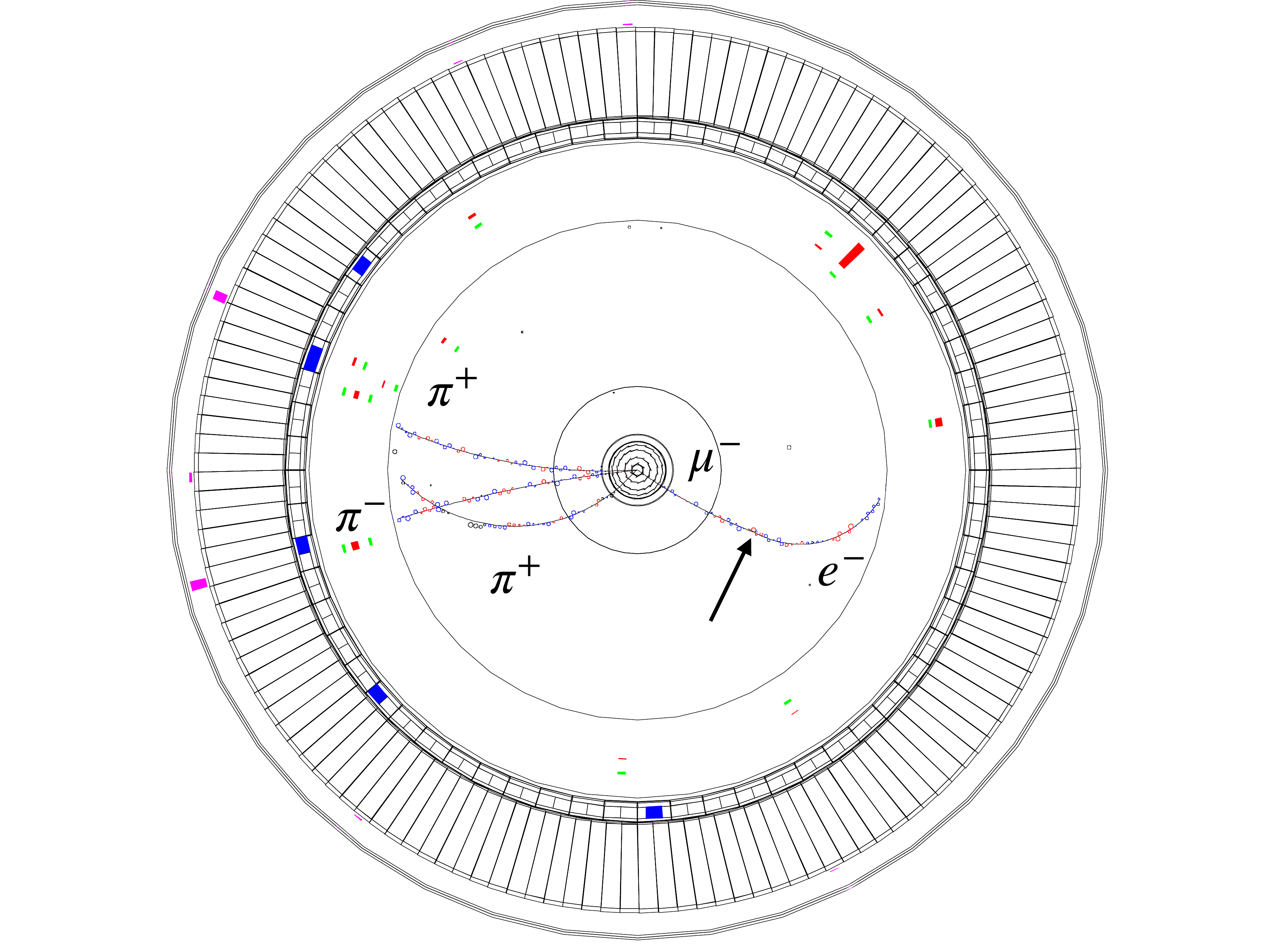}
\caption{Event display of a MC event $\ee\to\tat\to(\pi^+\pi^-\pi^+\bar{\nu}_\tau)(\mu^-\bar\nu_\mu\nu_\tau)$ with $\menn$ decay in the CDC (the arrow points to the decay vertex). The Belle detector, without the KLM, is shown projected onto the $x$--$y$ plane.}
\label{fig:1} 
\end{figure}

The Belle reconstruction algorithm can find separately the trajectories of the parent muon and the daughter electron from the $\menn$ decay: together these appear as a kinked track (kink) in the CDC. These trajectories can be fitted to the point of their closest approach to obtain the muon decay vertex. Figure~\ref{fig:1} shows the event display of a typical signal event generated with the MC simulation, where a kink from the $\menn$ decay is clearly observed.

We perform $\cascade$ event selection based on the $\tat$-pair event topology, kink pattern, and $\menn$ decay features. Here we provide only a qualitative discussion, while a detailed description of the selection can be found in Ref.~\cite{Belle:2023dyc}. We start with a standard preselection of $\ee\to\tat$ events with $\tmnn$ decay on the signal side. We are not interested in a particular decay mode of the tagging $\tau$ lepton; therefore, we content ourselves only with its decay topology. As the majority of the $\tau$ decay modes have one or three charged particles in the final state with several $\pi^0$s and photons, we require the event topology to be 1--1 or \mbox{1--3}: one charged track originating from the beam interaction point (IP) in the signal hemisphere (muon candidate from the $\tmnn$ decay) and one or three charged tracks from the IP in the tag hemisphere. In the signal hemisphere, we additionally require one track displaced from IP, which is then used as an electron candidate from the $\menn$ decay.

In the first step, we focus on the suppression of the non-$\tat$ background, such as hadron and muon production, two-photon processes, and Bhabha scattering. For this, we use the missing energy and momentum signature typical for $\tat$ pair events due to undetectable neutrinos in the final state. We additionally suppress Bhabha scattering by a loose electron veto in the tagging hemisphere using the standard Belle particle identification based on the information from the CDC, ACC, and ECL subdetectors~\cite{Nakano:2002jw}. 

We also suppress the background in the signal hemisphere from decay modes other than $\tmnn$, which mainly contain $\pi^-$ in the final state, often accompanied by $\pi^0$s. Thus, we veto events with significant energy deposits in the ECL associated with photons in the signal hemisphere.

The next step is the innovative part of the analysis since kinks have never been used in Belle studies before. The selection is based on two reconstructed tracks in the CDC: one originates in the vicinity of the IP (muon candidate) and ends inside the CDC volume, and the other track (electron candidate) is required to be inconsistent with being produced at the IP. Both trajectories are required to be located close to each other forming a kink, and the point of their closest approach is used as the muon decay vertex. Since kink trajectories are, on average, shorter than regular ones, we limit the number of CDC hits assigned to the tracks. We additionally suppress combinatorial background by vetoing signals in the Belle subsystems, which are absent for kink tracks (TOF, ECL, and KLM response for the muon track candidate and SVD and KLM response for the electron candidate).

\begin{figure}[htb]
  \includegraphics[width=1\linewidth]{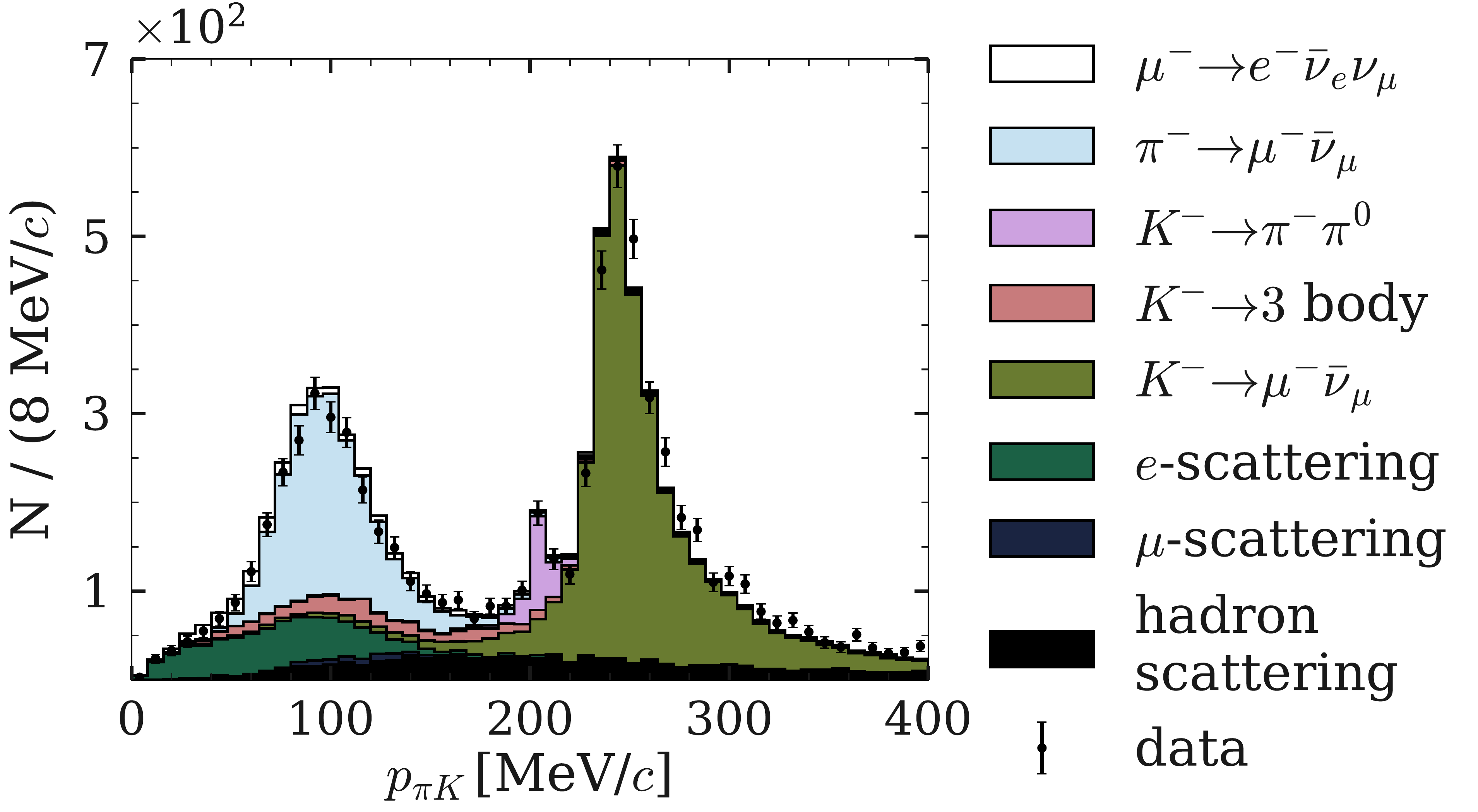}
\caption{Distribution of the momentum of the daughter, in the rest frame of the mother, for pion and kaon mass hypotheses assigned to the daughter and mother particles, respectively.}
\label{fig:2}
\end{figure}

After the application of the above two steps of selection, the nonkink background is suppressed to a negligible level, as confirmed using the MC simulation. The remaining events consist of the signal $\menn$ decay and real-kink background processes mimicking the signal: light meson decays ($\pimunu$, $\kmunu$, $\kpmunu$, $\kpenu$, $\kpipi$, $\kcpi$, and $\kpipipi$), and electron, muon, and hadron scattering. The $K^-$ and $\pi^-$ decays are mainly two-body and are characterized by monochromatic daughter particle momentum in the mother's rest frame when the correct pair of mass hypotheses is assigned to tracks. Figure~\ref{fig:2} shows such momentum distributions calculated for pion and kaon mass hypotheses assigned to the daughter and mother tracks, respectively.
The filled histograms correspond to MC events, while points with errors show the data. Peaks from the two-body decays are clearly observed: the narrowest one is from $\kpipi$ since the pair of mass hypotheses assigned to the tracks is correct for this decay. The dominant peak from the $\kmunu$ decay is slightly smeared by misinterpretation of the daughter mass hypothesis. Other backgrounds, which do not have such distinct characteristics, have a broad shape. Agreement between simulation and data is observed, and the background exceeds the signal by 2 orders of magnitude.

To suppress the remaining background with high efficiency, we employ a boosted decision tree classifier (BDT)~\cite{FREUND1997119,Hocker:2007ht}. We configure the BDT to distinguish signal from background based on 12 features selected by separation power and the physical characteristics of the signal and each background source. The first two features are the momentum of the daughter particle in the mother rest frame with two pairs of mass hypotheses: muon and pion ($p_{\mu\pi}$), and pion and kaon ($p_{\pi K}$) mass hypotheses assigned to the daughter and mother particles, respectively. The use of these variables is intended mainly to suppress the background from hadron kinks. 

The other five features are particle identification (PID) variables for the muon and electron candidates. The identification of the former is based only on the $dE/dx$ losses in the CDC; it includes the contradistinction of the muon hypothesis against electron, pion, kaon, and proton hypotheses. The electron candidate PID is based on the combination of the $dE/dx$ losses inside the CDC and the information from ECL. Here we only consider the electron hypothesis against the muon hypothesis. The PID variables are useful for the background suppression of all kink types. 

We also use parameters of the decay vertex in the BDT: the $z$ coordinate of the muon track endpoint and the distance between the kink tracks at the decay vertex. Their separation power is based on the difference in the momentum transfer, which depends on the kink type.

The last three features are based on the kinematics of the event and focused on the suppression of the residual background from non-$\tat$ processes, since about a third of hadron kinks originate from $\ee\to\qq$ events.

\begin{figure}[htb]
  \includegraphics[width=1\linewidth]{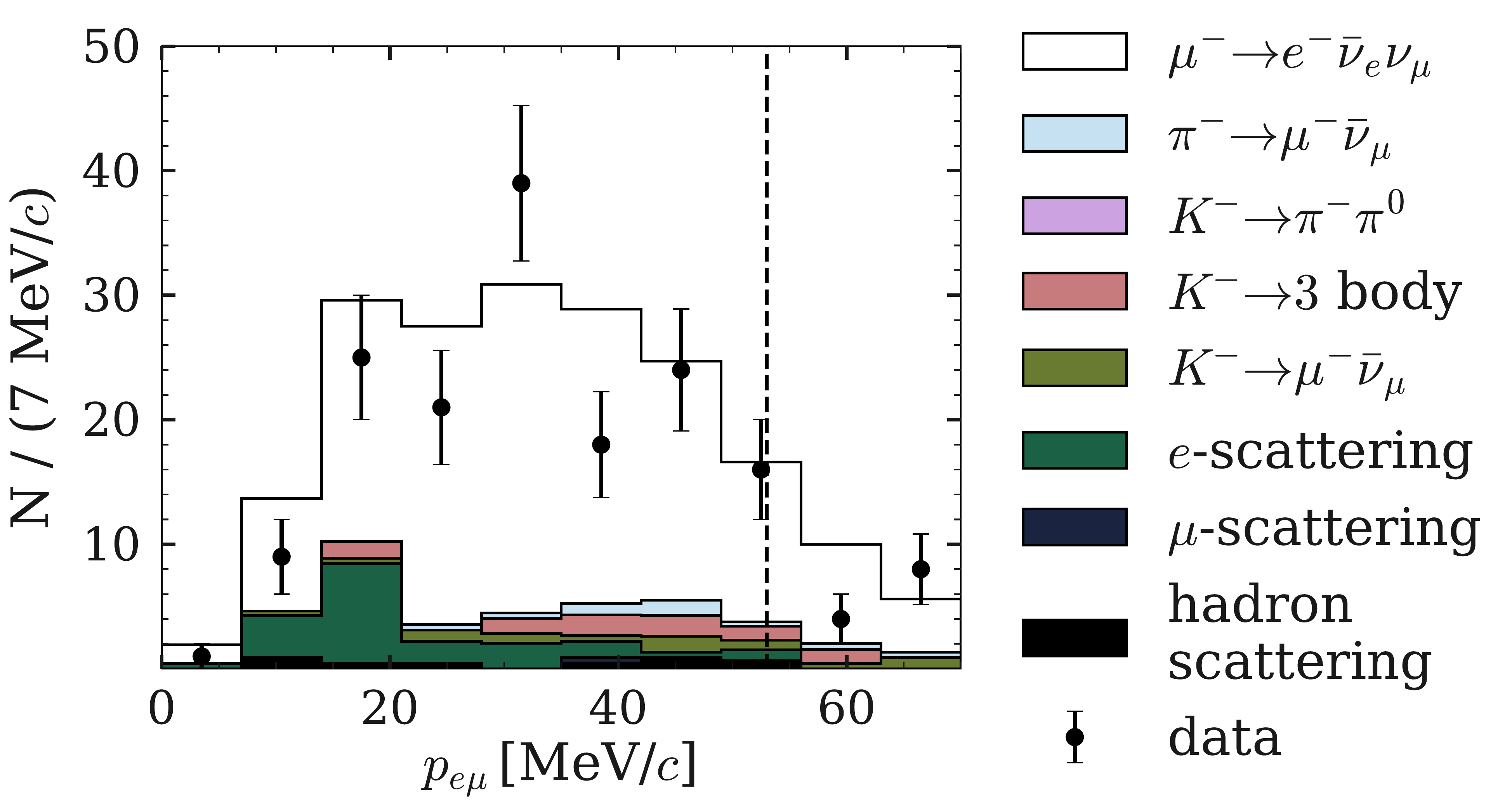}
\caption{Distribution of the momentum of the daughter, in the rest frame of the mother, for electron and muon mass hypotheses assigned to the daughter and mother particles, respectively. The dashed line shows the $53\,\text{MeV}/c$ threshold.}
\label{fig:3}
\end{figure}

The BDT application has a high signal efficiency (about $80\%$) while suppressing the background by a factor of 50. It is additionally checked with the MC simulation that no bias in the signal kinematic distributions related to $\xip$ is induced by the BDT selection. Figure~\ref{fig:3} shows the electron momentum in the muon rest frame with the corresponding mass hypotheses for the selected $\cascade$ events. The tail above the kinematic threshold of $53\,\text{MeV}/c$ in the $\menn$ decay is due to the imperfect resolution of the momentum of the mother and daughter particles, and the decay vertex. Finally, we obtain $165$ signal-candidate events in data, while based on the simulation, we estimate the number of signal events to be $139$ with $50$ background events. Although the number of events depends on the $\xip$ value, the effect is not significant for the reported study, and we estimate the number from the simulation with $\xip=0$.

To validate the MC simulation and estimate systematics, we use control samples with tagged kinks of different types: kaon two-body and three-body decays, $\pimunu$ decays, and hadron scattering selected from the $D^{*+}\to D^0(\to K^-\pi^+)\pi^+$ candidate sample, and electron scattering from the $\gamma$-conversion candidate sample. All samples of background-type kinks are larger than the corresponding background in the signal sample. The exploitation of the $D^{*+}$ and $\gamma$-conversion samples allows for the selection of specific types of kink without uncertainties induced by the BDT application. We also use $\kmunu$ and $\pimunu$ decays selected with BDT from the $\tau$ candidates sample. To take into account the BDT uncertainties, we compare light meson decay events from the $\tau$ and $D^{*+}$ samples. In general, agreement between the MC simulation and data is observed for all samples with minor discrepancies taken into account as systematics. 

To measure the Michel parameter $\xip$, we perform an unbinned maximum-likelihood fit to the $(y,\,\cos{\theta_{e}})\equiv(y,\,c)$ distribution, where $y$ is the electron energy in the muon rest frame divided by $m_\mu/2$, and ${\theta_{e}}$ is an angle of the electron emission direction in the muon rest frame. This angle carries information about the muon spin, taking into account its rotation in the magnetic field of the Belle detector. The definition of $\cos{\theta_{e}}$ and a detailed description of its calculation can be found in Ref.~\cite{Bodrov:2022mbd}. The probability density function (PDF) used in the fit is defined as follows:
\begin{eqnarray*}\label{eq:1}
\begin{aligned}
    \mathcal{P}(y,\,c;\, \xip)=p\,\mathcal{P}_\text{sig}(y,\,c;\, \xip) + (1-p)\, \mathcal{P}_\text{bckg}(y,\,c).
\end{aligned}
\end{eqnarray*}
Here $\mathcal{P}_\text{sig}(y,\,c;\, \xip)$ and $\mathcal{P}_\text{bckg}(y,\,c)$ are the signal and background PDF, respectively, and $p=0.74$ is the signal purity; we obtain all of them from the MC simulation. We calculate the signal PDF taking into account the reconstruction efficiency and detector resolution from two MC samples generated with the parameters $\xip=-1$ [$\mathcal{P}_{-1}(y,\,c)=\mathcal{P}_\text{sig}(y,\,c;\,-1)$] and $\xip=1$ [$\mathcal{P}_{+1}(y,\,c)=\mathcal{P}_\text{sig}(y,\,c;\,1)$] using the fact that the initial theoretical function is linear in the parameter $\xip$:
\begin{eqnarray*}\label{eq:2}
    \mathcal{P}_\text{sig}(y,\,c;\,\xip)\!&=&\!\dfrac{1}{2}\left\{\mathcal{P}_{+1}(y,\,c) + \mathcal{P}_{-1}(y,\,c)\right.\nonumber\\
    &&\left.+\xip\left[\mathcal{P}_{+1}(y,\,c) - \mathcal{P}_{-1}(y,\,c)\right]\right\}.
\end{eqnarray*}
The large size of the signal MC sample allows us to use $\mathcal{P}_\text{sig}(y,\,c;\, \xip)$ in the form of a $10\times 10$ histogram. By contrast, the background MC sample size is limited by the generating complexity; therefore, we choose $\mathcal{P}_\text{bckg}(y,\,c)$ as a smooth parametric function that describes the two-dimensional background histogram in $(y,\,c)$ with $\chi^2$ per degree of freedom $\approx 1$. This approach is tested using pseudoexperiments generated with 11 $\xip$ seed values from $-1$ to 1, and no bias is observed.

\begin{figure}[htb]
  \includegraphics[width=1\linewidth]{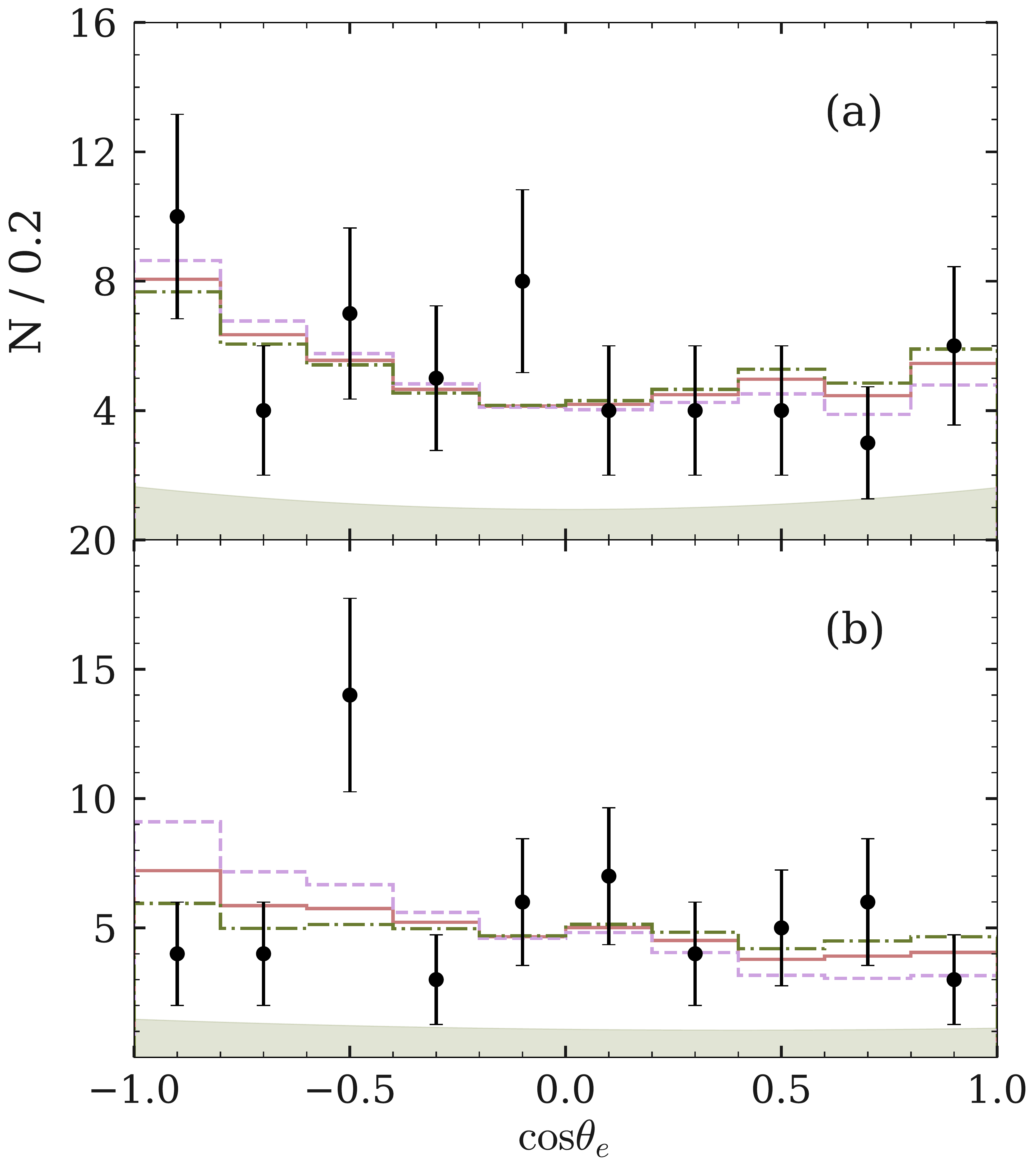}
\caption{Projection onto $\cos{\theta_e}$ for slices in $y$: (a)~$0.52 < y < 0.78$ and (b)~$0.78 < y < 1.3$. Points with error bars correspond to the data, the solid line corresponds to the $\xip=0.22$ fit function, the dashed line corresponds to the $\xip=-1$ fit function, the dash-dotted line corresponds to the $\xip=1$ fit function, and the shadowed area corresponds to the background function.}
\label{fig:4}
\end{figure}

Finally, the Michel parameter $\xip$ is found to be $\xip=0.22\pm0.94$ from the fit to the data. The result of the fit is shown in Fig.~\ref{fig:4} via projections onto $\cos{\theta_e}$ for two intervals in $y$ larger than $0.52$. A projection onto $\cos{\theta_e}$ for $y<0.52$ is not shown since there is almost no sensitivity to $\xip$ for this interval.

The systematic errors of the measurement are taken into account by assuming the most conservative approach. They are estimated in a blind manner using an ensemble of pseudoexperiments simulated for the whole physically allowed range of $\xip$ values from $-1$ to $1$. Each sample is generated according to the PDF, determined taking into account any discrepancies between the data and the MC simulation observed in the control kink samples. Four main sources of systematics are considered: the signal and background PDFs, particle identification, and the fit procedure. We estimate their contributions to be $0.14$, $0.20$, $0.24$, and $0.25$, respectively, leading to a total systematic uncertainty of $0.42$. It is additionally checked that systematics estimated from the data fit by variation of corresponding PDFs yields a lower value for each source compared to the blinded approach.

As a result, we measure the Michel parameter $\xip$ to be 
\begin{eqnarray*}
\xip=\res,
\end{eqnarray*}
where the first uncertainty is statistical and the second one is systematic. The obtained value is in agreement with the SM expectation of $\xip=1$ within the uncertainties. The precision of the reported $\xip$ measurement is significantly better than that obtained in the radiative $\tau$ decays~\cite{Shimizu:2017dpq}. 

In summary, we report the first direct measurement of the Michel parameter $\xip$ in the $\tmnn$ decay using the full data sample of $988\,\text{fb}^{-1}$ collected by the Belle detector. The obtained value $\xip=\res$ is in agreement with the SM prediction. The performed measurement proves the feasibility of the novel method proposed in Ref.~\cite{Bodrov:2022mbd}. This analysis demonstrated that the overwhelming background can be efficiently suppressed with the machine learning algorithm to a manageable level. Statistics is a limiting factor in this analysis, while systematics is smaller and remains well under control with various data-control samples. Thus, in future experiments, the systematics will not restrain the overall accuracy. In this study, one of the limiting factors for the precision turned out to be the kink-vertex resolution due to the absence of a dedicated kink reconstruction algorithm. While we could not overcome this problem in this analysis due to various nonphysical factors, it is feasible to implement such an algorithm in future experiments. Based on this experience, it should be possible to improve the result in the near future in the Belle~II experiment~\cite{Kou:2018nap}. Furthermore, the pioneering of this method opens a unique opportunity to measure the muon polarization in other weak processes like semileptonic decays of heavy mesons.

This work, based on data collected using the Belle detector, which was
operated until June 2010, was supported by 
the Ministry of Education, Culture, Sports, Science, and
Technology (MEXT) of Japan, the Japan Society for the 
Promotion of Science (JSPS), and the Tau-Lepton Physics 
Research Center of Nagoya University; 
the Australian Research Council including grants
DP210101900, 
DP210102831, 
DE220100462, 
LE210100098, 
LE230100085; 
Austrian Federal Ministry of Education, Science and Research (FWF) and
FWF Austrian Science Fund No.~P~31361-N36;
the National Natural Science Foundation of China under Contracts
No.~11675166,  
No.~11705209;  
No.~11975076;  
No.~12135005;  
No.~12175041;  
No.~12161141008; 
Key Research Program of Frontier Sciences, Chinese Academy of Sciences (CAS), Grant No.~QYZDJ-SSW-SLH011; 
Project ZR2022JQ02 supported by Shandong Provincial Natural Science Foundation;
the Ministry of Education, Youth and Sports of the Czech
Republic under Contract No.~LTT17020;
the Czech Science Foundation Grant No. 22-18469S;
Horizon 2020 ERC Advanced Grant No.~884719 and ERC Starting Grant No.~947006 ``InterLeptons'' (European Union);
the Carl Zeiss Foundation, the Deutsche Forschungsgemeinschaft, the
Excellence Cluster Universe, and the VolkswagenStiftung;
the Department of Atomic Energy (Project Identification No. RTI 4002) and the Department of Science and Technology of India; 
the Istituto Nazionale di Fisica Nucleare of Italy; 
National Research Foundation (NRF) of Korea Grant
Nos.~2016R1\-D1A1B\-02012900, 2018R1\-A2B\-3003643,
2018R1\-A6A1A\-06024970, RS\-2022\-00197659,
2019R1\-I1A3A\-01058933, 2021R1\-A6A1A\-03043957,
2021R1\-F1A\-1060423, 2021R1\-F1A\-1064008, 2022R1\-A2C\-1003993;
Radiation Science Research Institute, Foreign Large-size Research Facility Application Supporting project, the Global Science Experimental Data Hub Center of the Korea Institute of Science and Technology Information and KREONET/GLORIAD;
the Polish Ministry of Science and Higher Education and 
the National Science Center;
the Ministry of Science and Higher Education of the Russian Federation, Agreement 14.W03.31.0026, 
and the HSE University Basic Research Program, Moscow; 
University of Tabuk research grants
S-1440-0321, S-0256-1438, and S-0280-1439 (Saudi Arabia);
the Slovenian Research Agency Grant Nos. J1-9124 and P1-0135;
Ikerbasque, Basque Foundation for Science, Spain;
the Swiss National Science Foundation; 
the Ministry of Education and the Ministry of Science and Technology of Taiwan;
and the United States Department of Energy and the National Science Foundation.
These acknowledgements are not to be interpreted as an endorsement of any
statement made by any of our institutes, funding agencies, governments, or
their representatives.
We thank the KEKB group for the excellent operation of the
accelerator; the KEK cryogenics group for the efficient
operation of the solenoid; and the KEK computer group and the Pacific Northwest National
Laboratory (PNNL) Environmental Molecular Sciences Laboratory (EMSL)
computing group for strong computing support; and the National
Institute of Informatics, and Science Information NETwork 6 (SINET6) for
valuable network support.

\bibliography{bibl}

\end{document}